\documentclass{IEEEtran}
\pdfoutput=1
\usepackage{amsmath}
\usepackage{amssymb}
\usepackage{cite}
\usepackage{graphicx}
\usepackage{subfig}
\usepackage[nolist]{acronym}
\usepackage{booktabs}
\usepackage{comment}
\usepackage{flushend}
\usepackage{paralist}
\usepackage{color}
\usepackage{soul}
\usepackage{tikz}
\usepackage{pgfplots}
\usepackage{bm}
\pgfplotsset{compat=newest}

\usepackage{anyfontsize}

\newcommand{\ve}[1]{\mathbf{#1}}  					
\newcommand{\ma}[1]{\mathbf{#1}}  					
\newcommand{\mat}[1]{\mathbf{#1}}					
\renewcommand{\H}{^{\mathrm{H}}}
\newcommand{\T}{^{\mathrm{T}}}

\newcommand{\vectorize}[1]{\operatorname{vec}\!\left(#1\right)}

\renewcommand{\mod}[2]{\left<#1\right>_{#2}}

\newcommand{\secref}[1]{Section~\ref{#1}}

 \newcommand{\MMa}[1]{#1} 

\begin{document}
\title{GFDM Transceiver using Precoded Data and Low-complexity Multiplication in Time Domain}
\author{
	\IEEEauthorblockN{
		\normalsize{
            Ivan Gaspar\IEEEauthorrefmark{1},
			Maximilian~Matth\'e\IEEEauthorrefmark{1},
            Nicola Michailow\IEEEauthorrefmark{1},
			Luciano~Leonel~Mendes\IEEEauthorrefmark{2},
            Dan Zhang\IEEEauthorrefmark{1},
            Gerhard~Fettweis\IEEEauthorrefmark{1},
		}
	}\\
	\IEEEauthorblockA{
		\small{
			\IEEEauthorrefmark{1}Vodafone Chair Mobile Communication Systems, Technische Universit\"at Dresden, Germany\\
			\IEEEauthorrefmark{2}Instituto Nacional de Telecomunica\c{c}\~{o}es, Sta. Rita do Sapuca\'{\i}, MG, Brazil\\
			\texttt{\{first name.last name\}@ifn.et.tu-dresden.de}
		}
	}
\thanks{This work is supported by 5GNOW (Project 318555 FP7-ICT) and CNPq, Conselho Nacional de Desenvolvimento Cient\'{\i}fico e Tecnol\'{o}gico - Brasil.}
\thanks{Digital Object Identifier ...}
}
\maketitle


\begin{acronym}
  \acro{3G}{third generation}
  \acro{4G}{fourth generation}
  \acro{5G}{Fifth generation}
  \acro{ASIP}{Application Specific Integrated Processors}
  \acro{AWGN}{additive white Gaussian noise}
  \acro{BS}{base station}
  \acro{CP}{cyclic prefix}
  \acro{CR}{Cognitive Radio}
  \acro{CSMA}{carrier sense multiple access}
  \acro{DFT}{discrete Fourier transform}
  \acro{EPC}{evolved packet core}
  \acro{FBMC}{Filterbank multicarrier}
  \acro{FDMA}{frequency division multiple access}
  \acro{FPGA}{Field Programmable Gate Array}
  \acro{FTN}{Faster than Nyquist}
  \acro{FT}{Fourier transform}
  \acro{GFDM}{Generalized Frequency Division Multiplexing}
  \acro{ICI}{intercarrier interference}
  \acro{IDFT}{Inverse Discrete Fourier Transform}
  \acro{IMS}{IP multimedia subsystem}
  \acro{IoT}{Internet of Things}
  \acro{IP}{Internet Protocol}
  \acro{ISI}{intersymbol interference}
  \acro{IUI}{inter-user interference}
  \acro{LTE}{Long-Term evolution}
  \acro{M2M}{Machine-to-Machine}
  \acro{MA}{multiple access}
  \acro{MF}{Matched filter}
  \acro{MMSE} {minimum mean square error}
  \acro{MSE}{mean-squared error}
  \acro{NFV}{network functions virtualization}
  \acro{OFDM}{Orthogonal Frequency Division Multiplexing}
  \acro{OOB} {out-of-band}
  \acro{OQAM} {Offset Quadrature Amplitude Modulation} 
  \acro{PAPR}{peak to average power ratio}
  \acro{PHY}{physical layer}
  \acro{RC}{raised cosine}
  \acro{SC-FDE}{Single Carrier Frequency Domain Equalization}
  \acro{SC-FDMA}{Single Carrier Frequency Domain Multiple Access}
  \acro{SDN}{software-defined network}
  \acro{SDR}{software-defined radio}
  \acro{SDW}{software-defined waveform}
  \acro{SER}{symbol error rate}
  \acro{SIC} {successive interference cancellation}
  \acro{V-OFDM}{Vector OFDM}
  \acro{ZF}{zero-forcing}
  \acro{WLAN}{wireless Local Area Network}
  \acro{WRAN}{Wireless Regional Area Network}
  \acro{STFT}{short-time Fourier transform}
\end{acronym}

\begin{abstract}
Future wireless communication systems are demanding a more flexible physical layer. GFDM is a block filtered multicarrier modulation scheme proposed  to add multiple degrees of freedom and cover other waveforms in a single framework. In this paper, GFDM modulation and demodulation will be presented as a frequency-domain circular convolution, allowing for a reduction of the implementation complexity when MF, ZF and MMSE filters are employed in the inner and outer receiver operation. Also, precoding is introduced to further increase GFDM flexibility, addressing a wider set of applications.
\end{abstract}


\section{Introduction}\label{sec:1}

\ac{5G} networks are requiring a new level of flexibility on the \ac{PHY}. Although higher throughput keeps pushing the spectral efficiency to higher standards, new services will also demand very low latency, massive capacity for multiple connections and very low power consumption. The \ac{PHY} must be flexible to address several different scenarios.

\ac{GFDM} \cite{Gaspar2013} is a recent waveform that can be engineered to address various use cases. \ac{GFDM} arranges the data symbols in a time-frequency grid, consisting of $M$ subsymbols and $K$ subcarriers, and applies a circular prototype filter for each subcarrier. \ac{GFDM} can be easily configured to cover other waveforms, such as \ac{OFDM} and \ac{SC-FDMA} as corner cases. The subcarrier filtering can reduce the \ac{OOB} emissions, control \ac{PAPR} and allow dynamic spectrum allocation. \ac{GFDM}, with its block-based structure, can reuse several solutions developed for \ac{OFDM}, for instance, the concept of a \ac{CP} to avoid inter-frame interference. Hence, frequency-domain equalization can be efficiently employed to combat the effects of multipath channels prior to the demodulation process. With these features, \ac{GFDM} can address the requirements of 5G networks.

The main disadvantage of a flexible waveform is the complexity required for its implementation. Reducing the \ac{GFDM} complexity is essential to bring its flexibility to \ac{5G} networks.
The understanding of GFDM as a filtered multicarrier scheme leads to a modem implementation based on time-domain circular convolution for each subcarrier, performed in the frequency-domain \cite{Gaspar2013}. This paper follows the well known polyphase implementation of filter bank modulation \cite{farhang2011ofdm,renfors14,Banelli2014} and applies it to GFDM. 
	
As one main contribution, the investigation in this paper reveals that the Poisson summation formula~\cite{Benedetto97} can be utilized in the demodulation process, considering a per-subsymbol circular convolution and decimation in the frequency-domain. This operation can be performed as an element-wise multiplication with subsequent simple $M$-fold accumulation in the time-domain. This simple re-orientation of the data symbol processing allows for a considerable reduction in complexity, which makes \ac{GFDM} applicable in a wide range of scenarios. 

This paper also contributes to reformulate the modulation and demodulation as matrix-vector operations based on $M(K)$-point \acp{DFT}. As an outcome, it becomes evident that \ac{DFT} can actually be considered as a precoding operation. Moreover, they can be replaced by other transformations as well.  This observation facilitates arbitrary precoding of the GFDM data, which adds a new level of flexibility to the system.

It is worth mentioning that the proposed low complexity signal processing complements the work in \cite{Gaspar2013} because the requirement for block alignment is loosened. It can be usable for supporting pipeline inner receiver implementation, particularly when building synchronization and channel estimation circuits for embedded training sequences. It can also be beneficial to develop future non-linear and recursive detection algorithms.



\section{Classical GFDM Description and Low-Complexity Reformulation}\label{sec:2}

Consider a wireless communication system that transmits data in a block-based structure that consists of $K$ subcarriers and $M$ subsymbols. Let $N=KM$ be the total number of data symbols in the block and $d_{k,m}$, $k=0, \dots, K-1$, \mbox{$m=0, \dots, M-1$}, denote the complex valued data symbol that is transmitted on the $k$th subcarrier and $m$th subsymbol. The classical description of \ac{GFDM} signal generation \cite{Gaspar2013} is given by
\begin{align}\label{eq:classicTx}
  x[n] &= \sum_{k=0}^{K-1}\left(g[n] \exp(j2\pi \tfrac{k}{K}n)\right)\circledast d_k[n]\\
\text{with } d_k[n] &= \sum_{m=0}^{M-1}d_{k,m}\delta[n-mK],
\end{align}
\def\DFT{\mathcal{F}}

\noindent where $n=0,\dots,N-1$, $\circledast$ describes circular convolution carried out with period $N$ and $g[n]$ denotes the impulse response of the transmit prototype filter. \\
%
\indent Using the $N$-point \ac{DFT} $\DFT_N$, \eqref{eq:classicTx} can be carried out as
\begin{align}\label{eq:modInFD}
x[n] &=\DFT_N^{-1}
\sum_{k=0}^{K-1}\{
     \underbrace{\DFT_N\{g[n]\exp(j2\pi \tfrac{k}{K}n)\}}_{G[\mod{f-kM}{N}]}
     \underbrace{\DFT_N\{d_k[n]\}}_{\DFT_{M}\{d_{k,(\cdot)}\}|_f}\},
\end{align}
where $\mod{\bullet}{N}$ denotes the remainder modulo $N$ and
\mbox{$G[f]=\DFT_{N}\{g[n]\}$}.  Due to periodicity of $\DFT_M$
with period $M$ and $f=0,\dots,MK-1$, $\DFT_M\{d_{k,(\cdot)}\}$ is
concatenated $K$ times.  This leads to the low-complexity modulator
implementation described in \cite{Gaspar2013}.  Now,
expressing the convolution in (\ref{eq:classicTx}) explicitly, the
transmission equation can be reformulated as \cite{Banelli2014}
\begin{align}\label{eq:modInTD}
  x[n] &=
	\sum_{m=0}^{M-1}g[\mod{n-mK}{N}]\underbrace{\sum_{k=0}^{K-1}d_{k,m}\exp(j2\pi \tfrac{k}{K}n)}_{K\DFT^{-1}_K\{d_{(\cdot),m}\}|_n}.
\end{align}
The sequence obtained from the \ac{IDFT} is concatenated $M$
times in time-domain due to $n=0,\dots,MK-1$.  Such multiplication in
the time-domain expresses the convolution of the subcarrier filter
with the data in the frequency-domain for each subsymbol.\\
\indent Assuming a flat and noiseless channel, the convolution of the received signal $y[n]=x[n]$ with a demodulation filter $\gamma[n]$, is given
by
\begin{align}\label{eq:classicRX}
\hat{d}_{k,m}&=\left(\gamma^*[\mod{-n}{N}]\circledast y[n]\exp(-j2\pi \tfrac{k}{K}n)\right)|_{n=mK}.
\end{align}
With the help of \ac{DFT}, \eqref{eq:classicRX} is
written as
\begin{small}
\begin{align}\label{eq:demodInFD}
  \hat{d}_{k,m}&=\DFT^{-1}_N\{
     \underbrace{\DFT_N\{\gamma^*[\mod{-n}{N}]\}}_{\Gamma^*[f]}
     \underbrace{\DFT_N\{y[n]\exp(-j2\pi \tfrac{k}{K}n)\}}_{Y[\mod{f+kM}{N}]}
   \}|_{mK},
\end{align}
\end{small}

\noindent where $\DFT_{N}\{\gamma[n]\}=\Gamma[f]$
and $\DFT_N\{y[n]\}=Y[f]$. For example, this leads to the low-complexity demodulator
implementation described in \cite{Gaspar2013} for the \ac{MF} $\gamma[n]=g[n]$, which was later extended to more general
filter types in \cite{Matthe2014}.
Now, directly expressing the convolution and sampling in
\eqref{eq:classicRX} leads to the time-domain multiplication
\begin{equation}\label{eq:demodInTd}
\begin{split}
  \hat{d}_{k,m}&=\sum_{n=0}^{N-1}\gamma^*[\mod{n-mK}{N}]y[n]\exp(-j2\pi
  \tfrac{kM}{KM}n)\\
&= \DFT_N\{\gamma^*[\mod{n-mK}{N}]y[n]\}|_{kM},
\end{split}
\end{equation}
where the $N$-point DFT is only evaluated at every $M$th sample.
According to
the Poisson summation formula \cite{Benedetto97}, this can be reduced to a $\tfrac{N}{M}=K$ point \ac{DFT}
\begin{align}\label{eq:demodInTdPoisson}
  \DFT_N\{u[n]\}|_{kM}&=\DFT_{\tfrac{N}{M}}\left.\left\{\sum_{m=0}^{M-1}u[\mod{n-mK}{N}]\right\}\right|_k.
\end{align}

Eq. \eqref{eq:demodInTd} describes a sampled \ac{STFT}.  This is
clear since the GFDM demodulator is actually performing a Gabor
transform \cite{Matthe2014}, which can be understood as a sampled
\ac{STFT}. For this reason, (\ref{eq:demodInTd}) does not allow to
employ different prototype filters per subcarrier.
\MMa{This constraint is circumvented by performing equalization before
demodulation or the presented scheme can be used to initially access
the channel state information by demodulation of training data.}
Also, the proposed scheme facilitates the implementation of odd number of subsymbols \cite{Matthe2014} avoiding the use of $M$-point \ac{DFT} and \ac{IDFT} required in \cite{Gaspar2013}.\\
\indent Notice that (\ref{eq:demodInTd}) holds when \ac{MF}, \ac{ZF} or \ac{MMSE} filters are used. \ac{ZF} or \ac{MMSE} filters can deliver good \ac{SER} performance, while \ac{MF} needs \ac{SIC} to remove the \ac{SER} floor caused by the self-interference. Hence, (\ref{eq:demodInTd}) shall be used to implement \ac{ZF} and \ac{MMSE} demodulators, while \ac{MF} can be considered when the transmit filters are orthogonal or \ac{OQAM} is employed.


\section{GFDM Matrix Model}\label{sec:3}
\begin{figure}
	\centering
	\includegraphics[scale=.75]{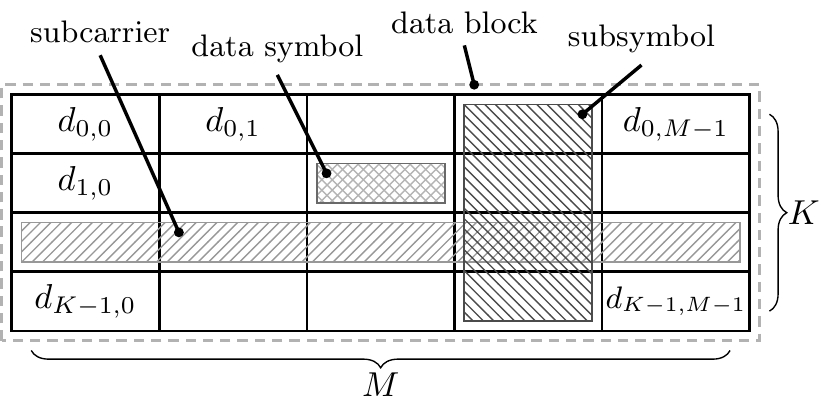}
	\caption{Overview of block structure and terminology.}
	\label{fig-terminology}
\end{figure}

\begin{figure*}[t]
  \centering
  \includegraphics[scale=.85]{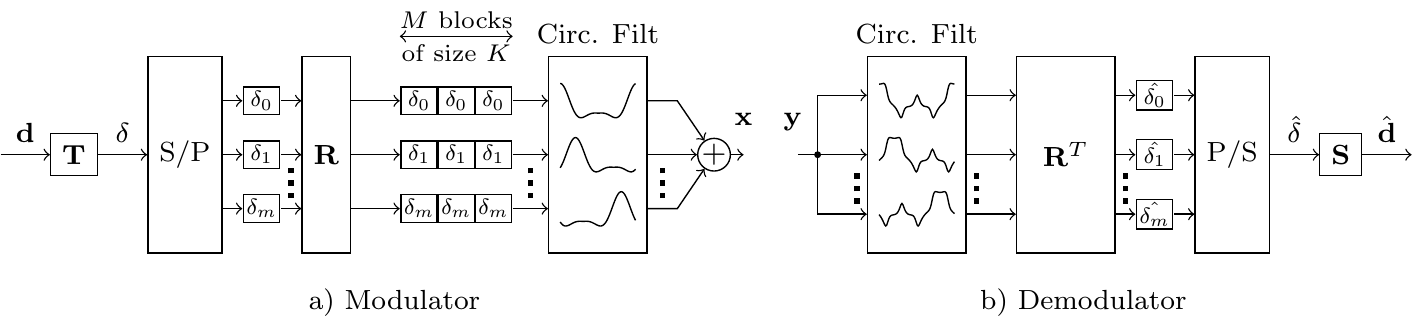}
  \caption{Block diagram of low-complexity modulator and demodulator with precoding.}
  \label{fig:blockDiag}
\end{figure*}
Arranging the data symbols $d_{k,m}$ in a two-dimensional structure leads to the data matrix $
  \mat{D}=
  \begin{pmatrix}
    \mathbf{d}_{0} \,\, \mathbf{d}_{1} \,\, \cdots \,\, \mathbf{d}_{M-1}
  \end{pmatrix},
$
where $\ve{d}_m = \begin{pmatrix} d_{0,m} & d_{1,m} & \dots & d_{K-1,m} \end{pmatrix}\T$ denotes the data symbols transmitted in the $m$th subsymbol.\\
\indent According to the \ac{GFDM} description in \secref{sec:2}, the rows and columns correspond to the time and frequency resources in Fig. \ref{fig-terminology}, respectively. Hence, the $k$th row of $\mathbf{D}$ represents the data symbols transmitted in the $k$th subcarrier, $\ve{d}_k = \begin{pmatrix} d_{k,0} & d_{k,1} & \dots & d_{k,M-1} \end{pmatrix}\T$.
The time-domain circular convolution of the modulation process can be expressed in the frequency-domain as \cite{Gaspar2013}
\begin{equation}
\ve{x} = \frac{1}{N}\ma{W}_N\H \sum\limits_{k=0}^{K-1} \ma{P}^{(k)} \ma{G} \ma{R}^{(K,M)} \ma{W}_M \ve{d}_k.
\label{ifftmattx}
\end{equation}
\indent For each subcarrier, the data vector is taken to the frequency-domain by the $M$-point \ac{DFT} matrix $\ma{W}_M$. The corresponding time-domain up sampling operation is realized in the frequency-domain by duplicating the transformed data symbols vectors $K$ times, using the repetition matrix $\ma{R}^{(K,M)} = \mathbf{1}_{K,1}\otimes \ma{I}_M$, where $\ma{I}_M$ is an $M$ size identity matrix, $\mathbf{1}_{i,j}$ is a $i\times j$ matrix of ones and $\otimes$ is the Kronecker product. Each subcarrier is then filtered with $\ma{G}=\text{diag}(\ma{W}_{N}\ve{g})$, where $\text{diag}(\bullet)$ returns a matrix that contains the argument vector on its diagonal and zeros otherwise and $\ve{g}$ is the vector containing the transmit filter impulse response. An up-conversion of the $k$th subcarrier to its respective subcarrier frequency is performed by the shift matrix
\begin{equation}
  \mathbf{P}^{(k)}=\mathbf{\Psi}\left(\mathbf{p}^{(k)}\right)\otimes\ma{I}_M,
\end{equation}
where $\mathbf{\Psi}(\bullet)$ returns the circulant matrix based on the input vector and $\mathbf{p}^{(k)}$ is a column vector where the $k$th element is 1 and all others are zero. The $K$ subcarriers are summed and transformed back to the time-domain with $\ma{W}_N^H$ to compose the GFDM signal. On the demodulator side, the recovered data symbols for the $k$th subcarrier are given by
\begin{equation}
\hat{\ve{d}}_k = \frac{1}{M}\ma{W}_{\!\!M\,\,}\H \left(\ma{R}^{(K,M)}\right)\T \ma{\Gamma} \left(\ma{P}^{(k)} \right) \T \ma{W}_{N} \ve{y},
\end{equation}
where $\mathbf{y}$ is the equalized vector at the input of the demodulator, \mbox{$\bm{\Gamma}=\text{diag}(\mat{W}_{N}\bm{\gamma})$} with $\bm{\gamma}$ being the demodulation filter impulse response, e.g., \ac{MF}, \ac{ZF} or \ac{MMSE} filters.\\
\indent The modulation and demodulation processes can be simplified just by changing the processing order of the data symbols, as derived in (\ref{eq:modInTD}). In this new matrix model, the GFDM vector is given by
\begin{equation}
\ve{x} = \sum\limits_{m=0}^{M-1} \ma{P}^{(m)} \text{diag}(\bm{g}) \ma{R}^{(M,K)} \ma{W}_K^H \ve{d}_m.
\label{noifftmattx}
\end{equation}
Observe that (\ref{noifftmattx}) is similar to a polyphase filter structure, but with the difference that cyclic time-shifts are used here and frequency-domain convolution is performed in time-domain as element-wise vector multiplication. With this approach, the first step is to obtain a time-domain version of $ \ve{d}_m $ by multiplying it with an inverse \ac{DFT} (IDFT) matrix $\ma{W}_K\H$. $M$ times upsampling in frequency-domain is performed in time by duplicating the transformed data symbols with a repetition matrix $\ma{R}^{(M,K)}$. Each subsymbol is then pulse-shaped with $\bm{g}$. $\ma{P}^{(m)}$ shifts the $m$th subsymbol to its respective position in time. The GFDM signal is obtained by summing all the pulse-shaped subsymbols, with no need of the $N$-IDFT domain conversion in (\ref{ifftmattx}).

The demodulation operations are also simplified by using the circular-convolution in frequency-domain, leading to
\begin{equation}
\hat{\ve{d}}_m = {\mat{W}_{K}}\left(\ma{R}^{(M,K)}\right)\T \text{diag}(\bm{\gamma}) \left(\ma{P}^{(m)} \right) \T \ve{y}.
\label{detec-nofft}
\end{equation}
\indent While the new approach describing \ac{GFDM} modulation and demodulation as convolution in the frequency-domain reduces the implementation complexity, in the next section this principle will be expanded to a more general structure, where domain conversions will be understood as a precoding process.

\begin{figure*}
	\centering
	\includegraphics[scale=1]{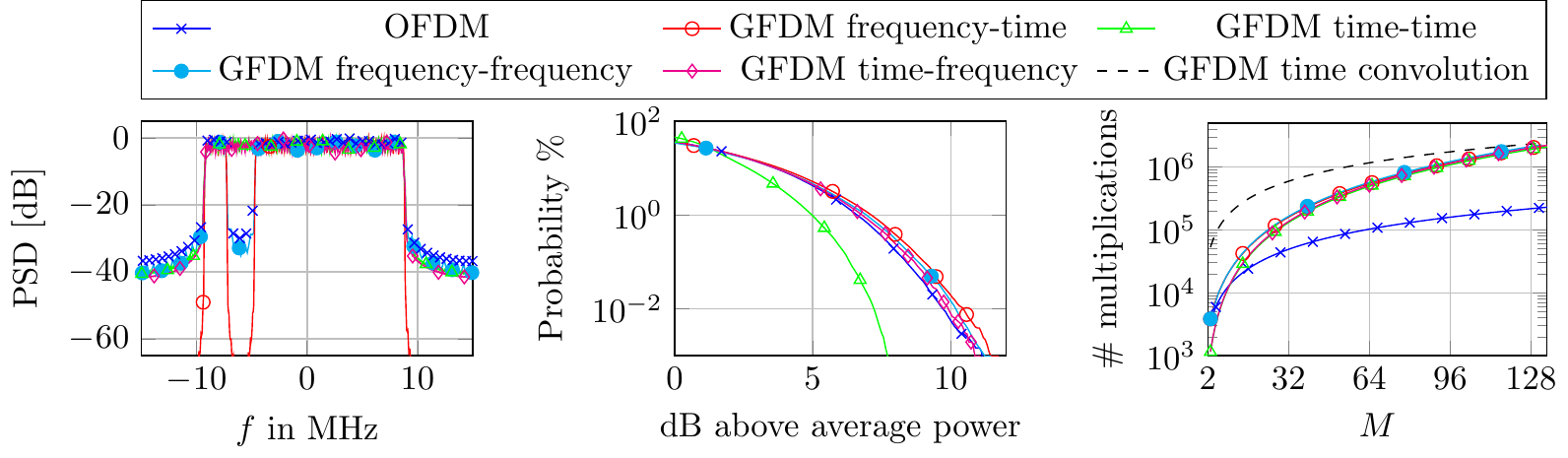}
	\captionsetup{justification=centering}
	\caption{OOB, PAPR and complexity analysis for precoded GFDM.\\
	FT/FF: $M=16$, $M_\text{on}=15$, $K=128$, $K_\text{on}=75$, TT/TF: $K=1$, $M=1200$, OFDM: $N=2048$, $N_\text{on}=1200$.}
	\label{fig:oob_papr_cmplx}
\end{figure*}

\section{Expanding GFDM features with precoding}
%
Eqs. (\ref{noifftmattx}) and (\ref{detec-nofft}) bring a new interpretation of the GFDM chain that further increases the flexibility of this waveform. Let the transmit vector from (\ref{noifftmattx}) be redefined as
\begin{equation}
\ve{x} = \sum\limits_{m=0}^{M-1} \ma{P}^{(m)} \text{diag}(\bm{g}) \ma{R}^{(M,K)} \bm{\delta}_m,
\label{precoding-tx}
\end{equation}
\noindent where
\begin{equation}
  \bm{\delta}_m=\ma{T}\ve{d}_m
  \label{eq-precoding-vector} 
\end{equation}
\noindent is the information coefficient vector to be transmitted and $\ma{T}$  is a transformation matrix. For classical \ac{GFDM}, where a time-frequency grid is used to transmit the information, $\ma{T}=\ma{W}_K^H$. However, (\ref{eq-precoding-vector}) can be seen as precoding the information vector by a generic matrix $\ma{T}$. Different matrices can be used to achieve different requirements. Hence, by transmitting precoded data, GFDM can be even more flexible to address requirements of future wireless networks.\\
\indent On the receiver side, the recovered coefficients are given by
\begin{equation}
\hat{\bm{\delta}}_m = \left(\ma{R}^{(M,K)}\right)\T \text{diag}(\bm{\gamma}) \left(\ma{P}^{(m)} \right) \T \ve{y}.
\label{precoding-rx}
\end{equation}
\noindent The estimated data symbols are obtained by reverting the precoding operation on (\ref{precoding-rx}), leading to
\begin{equation}
  \hat{\ve{d}}_m=\mathbf{S}\hat{\bm{\delta}}_m.
\end{equation}
\noindent where $\mat{S}$ removes precoding. For conventional GFDM, \mbox{$\mat{S}=\mat{W}_{K}$}. A block diagram that described the proposed GFDM modem is presented in Fig. \ref{fig:blockDiag}. The demodulation part can be implemented to run continuously, which can be used for tracking specific properties of the estimated information coefficient vector $\hat{\bm{\delta}}_m$ containing training data.\\
\indent The precoding concept can be enlarged to embrace any transform taking into account the two dimensions of the \ac{GFDM} data block $\ma{D}$, which is a bi-dimensional structure defined as a frequency-time grid spanning over several subcarriers and subsymbols, as depicted in Fig. \ref{fig-terminology}. Hence, in the default configuration, the columns of the matrix contain data defined in the frequency-domain, while the rows have data in the time-domain. The precoded data block $\bm{\Delta}$ can be generalized to
\def\vp{\vphantom{\ma{T}_\text{c}}}
\begin{equation}
	\underset{K \times M}{\bm{\Delta}\vp} = \underset{K \times K}{\ma{T}_\text{c}} \cdot \underset{K \times M}{\ma{D}\vp} \cdot \underset{M \times M}{\ma{T}_\text{r}},
	\label{eq-precoding-matrix}
\end{equation}
where $\ma{D}$ is now a data block defined in an arbitrary domain, $\ma{T}_\text{c}$ is a $K$ by $K$ precoding matrix applied to the columns and $\ma{T}_\text{r}$ is a $M$ by $M$ precoding matrix applied to the rows of $\ma{D}$. For instance, choosing $\ma{T}_\text{c}$ and $\ma{T}_\text{r}$ to be Fourier matrices yields four domains for the \ac{GFDM} data block,\MMa{ as given in Tab.~\ref{tab:precoding} , where the domain frequency-time (FT) corresponds to standard \ac{GFDM}.} 
\def\vp2{\vphantom{\sum_{1}^{1}}}
\begin{table}[h]
	\centering
	\caption{Precoding in time (T) and frequency (F) domains}
	\begin{tabular}{l l l}
		\hline
		\textbf{Domain} & \textbf{Operation} & \textbf{Number of multiplications} \\
		\hline
		\text{FT}		& $\vp2\bm{\Delta}	= \ma{W}_K\H \ma{D} \ma{I}_M$		& $MK\log_2(K)$\\
		\text{TT}		& $\vp2\bm{\Delta}	= \ma{I}_K   \ma{D} \ma{I}_M$		& 0\\
		\text{FF}		& $\vp2\bm{\Delta}	= \ma{W}_K\H \ma{D} \ma{W}_M\H$	& $MK\log_2(K)+KM\log_2(M)$\\
		\text{TF}		& $\vp2\bm{\Delta}	= \ma{I}_K   \ma{D} \ma{W}_M\H$		& $KM\log_2(M)$\\
		\hline
	\end{tabular}
	\label{tab:precoding}
\end{table}
The choice of the data domain has impact on the robustness of the system regarding time and frequency selective fading. The concept is
not limited to \ac{DFT} and, in general, any meaningful transform can be applied to the rows and columns.
%
%

\indent Taking the concept one step further, \eqref{eq-precoding-matrix} can be formulated as
\begin{equation}
	\vectorize{\bm{\Delta}} = \vectorize{\ma{T}_\text{c} \ma{D} \ma{T}_\text{r}}.
	\label{eq-vectorized}
\end{equation}
With the help of the Kronecker product, the right part of \eqref{eq-vectorized} can be expressed as
\begin{equation}
	\vectorize{\ma{T}_\text{c} \ma{D} \ma{T}_\text{r}} = \left( \ma{T}_\text{r}\T \otimes \ma{T}_\text{c} \right) \vectorize{\ma{D}}.
	\label{eq-identity}
\end{equation}
The resulting matrix
\begin{equation}
	\ma{\mathcal{T}} = \left( \ma{T}_\text{r}\T \otimes \ma{T}_\text{c} \right)
	\label{eq-generalT}
\end{equation}
can be seen as an even more general way of precoding that is not restricted to subcarriers or subsymbols but allows arbitrary coupling between any elements of $\ve{d}=\vectorize{\ma{D}}$. The precoding is then applied leading to $\bm{\delta} = \ma{\mathcal{T}} \ve{d}$.\\
%


\section{Out of Band, PAPR and Complexity Analysis}\label{sec:4}

\MMa{In particular, the FT mode allows for use of null subsymbols and subcarriers to support non-continuous bandwidth application, achieving a low \ac{OOB} radiation as shown in \mbox{Fig. \ref{fig:oob_papr_cmplx}}. Although this is not possible with the time-time (TT) case, where the system essentially becomes a single-carrier system occupying the whole bandwidth, this precoding is especially beneficial for power-limited systems that cannot afford complex precoding operations with a low \ac{PAPR}, as the \ac{PAPR} can be significantly reduced. In contrast, frequency-frequency precoding (FF) results in a \ac{DFT}-precoded \ac{GFDM} system, with a spectrum similar to OFDM. In FF mode, data is defined directly in the frequency domain, according to (\ref{eq:modInFD}), with one particular subsymbol only located on at most two frequency bins, assuming a transmit filter that is bandlimited to two subcarriers. This property simplifies equalization, as only ICI remains, but on the other hand reduces frequency diversity in channels with strong frequency-selectivity. The last (TF) case also corresponds to the single carrier, but with symbols spread in time, creating potential for exploiting time-diversity.}

The complexity of the precoding operation of the GFDM transceiver based on \ac{DFT} algorithms are evaluated in Table \ref{tab:precoding}. Number of complex valued multiplications, a costly operation in implementations, is the chosen criterion for the analysis. The comparison is carried out under the assumption that \mbox{$N=KM$} complex data symbols are transmitted. As the baseline for the comparison in Fig.\ref{fig:oob_papr_cmplx}, $MN$ multiplications from (\ref{precoding-rx}) are added to the precoding operations and the OFDM algorithm requires naturally the complexity of a \ac{DFT} operation, with $N\log_2(N)$ multiplications.
%
The GFDM modulator and demodulator proposed in \cite{Gaspar2013}
require the complexity of the $M$-point DFT algorithm with additional
$K$ times complex multiplications over repeated chunks of $M$ complex
samples, resulting from a \ac{DFT} operation. Even when the frequency
response of the demodulation filter is assumed to be sparse, e.g. when
roll-off is small and a \ac{ZF} filter span $L$ can be smaller than
the total number of subcarriers, the modem scheme presented in this
paper is less complex than the solution proposed in
\cite{Gaspar2013}.
%


\bibliographystyle{IEEEtran}
\bibliography{ref}

\begin{thebibliography}{1}
\providecommand{\url}[1]{#1}
\csname url@samestyle\endcsname
\providecommand{\newblock}{\relax}
\providecommand{\bibinfo}[2]{#2}
\providecommand{\BIBentrySTDinterwordspacing}{\spaceskip=0pt\relax}
\providecommand{\BIBentryALTinterwordstretchfactor}{4}
\providecommand{\BIBentryALTinterwordspacing}{\spaceskip=\fontdimen2\font plus
\BIBentryALTinterwordstretchfactor\fontdimen3\font minus
  \fontdimen4\font\relax}
\providecommand{\BIBforeignlanguage}[2]{{%
\expandafter\ifx\csname l@#1\endcsname\relax
\typeout{** WARNING: IEEEtran.bst: No hyphenation pattern has been}%
\typeout{** loaded for the language `#1'. Using the pattern for}%
\typeout{** the default language instead.}%
\else
\language=\csname l@#1\endcsname
\fi
#2}}
\providecommand{\BIBdecl}{\relax}
\BIBdecl

\bibitem{Gaspar2013}
I.~Gaspar \emph{et~al.}, ``{Low Complexity GFDM Receiver Based On Sparse
  Frequency Domain Processing},'' in \emph{IEEE 77th Vehicular Technology
  Conference (VTC Spring)}, June 2013.

\bibitem{farhang2011ofdm}
B.~Farhang-Boroujeny, ``{OFDM} versus filter bank multicarrier,'' \emph{Signal
  Processing Magazine, IEEE}, vol.~28, no.~3, pp. 92--112, 2011.

\bibitem{renfors14}
M.~Renfors, J.~Yli-Kaakinen, and F.~Harris, ``Analysis and design of efficient
  and flexible fast-convolution based multirate filter banks,'' \emph{Signal
  Processing, IEEE Transactions on}, vol.~62, no.~15, pp. 3768--3783, Aug 2014.

\bibitem{Banelli2014}
P.~Banelli, S.~Buzzi, G.~Colavolpe, A.~Modenini, F.~Rusek, and A.~Ugolini,
  ``{Modulation Formats and Waveforms for 5G Networks: Who Will Be the Heir of
  OFDM?: An overview of alternative modulation schemes for improved spectral
  efficiency},'' \emph{Signal Processing Magazine, IEEE}, vol.~31, no.~6, pp.
  80--93, Nov 2014.

\bibitem{Benedetto97}
J.~Benedetto and G.~Zimmermann, ``{Sampling multipliers and the Poisson
  Summation Formula},'' \emph{Journal of Fourier Analysis and Applications},
  vol.~3, no.~5, pp. 505--523, 1997.

\bibitem{Matthe2014}
M.~Matthe, L.~Mendes, and G.~Fettweis, ``{Generalized Frequency Division
  Multiplexing in a Gabor Transform Setting},'' \emph{Communications Letters,
  IEEE}, vol.~18, no.~8, pp. 1379--1382, Aug 2014.

\end{thebibliography}
\end{document}